\documentclass[twocolumn,showpacs,aps,prd,superscriptaddress]{revtex4}

\usepackage{graphicx}
\usepackage{dcolumn}
\usepackage{amsmath}
\usepackage{epsfig}

\input pubboard/babarsym

\def\BDstarDsstar{\ensuremath{  B^{0} \rightarrow {  D_s^{*+} D^{*-}}}\xspace}
\def\BDstarDs{\ensuremath{  B^{0} \rightarrow {   D_s^{+} D^{*-}}}\xspace}
\def\BDstarDss{\ensuremath{  B^{0} \rightarrow {  D_s^{(*)+} D^{*-}}}\xspace}

\def\inter{i}

\def\Dspi{{\ensuremath{\Dsp\pi^-}}\xspace}

\def\Dsstarpi{{\ensuremath{\Dspstar\pi^-}}\xspace}

\def\DsspiWS{{\ensuremath{\Dsps\pi^+}}\xspace}

\def\brphipi{\ensuremath{\BR(\dstophipi)}\xspace}

\def\mmiss{\ensuremath{M_{\rm miss}}\xspace}
\def\Dsp{{\ensuremath{ D_s^+}}\xspace}

\def\Dspstar{{\ensuremath{ D_s^{*+}}}\xspace}
\def\Dsps{{\ensuremath{ D_s^{(*)+}}}\xspace}

\def\Dsphipi{{ \ensuremath{ {D_s^{+}}\rightarrow \phi \pi^{+}}}\xspace}
\def\Dsgamma{{ \ensuremath{ {D_s^{*+}}\rightarrow {D_s^{+}} \gamma}}\xspace}

\def\dstophipi{{\ensuremath{\Ds\to \phi\pi^{+}}}\xspace}
\def\dstokstark{{\ensuremath{\Ds\to \Kstarzb K^+}}\xspace}
\def\dstoksk{{\ensuremath{\Ds\to \KS K^+}}\xspace}

\def\phikk{{\ensuremath{\phi\to { K^+K^-}}}}

\newcommand{\BaBarYear}       {03}
\newcommand{\BaBarNumber}     {001}
\newcommand{\SLACPubNumber} {9644}

\newcommand{\BaBarType}      {PUB}

\def\figurebox#1#2#3{%
    \def\arg{#3}%
     \ifx\arg\empty
    {\hfill\vbox{\hsize#2\hrule\hbox to #2{\vrule\hfill\vbox to #1{\hsize#2\vfill}\vrule}\hrule}\hfill}%
    \else
    {\hfill\epsfbox{#3}\hfill}%
    \fi}

\long\def\inst#1{\par\nobreak\kern 4pt\nobreak
    {\it #1}\par\vskip 10pt plus 3pt minus 3pt}

\begin{document}

\begin{flushleft}
\babar-\BaBarType-\BaBarYear/\BaBarNumber \\
SLAC-PUB-\SLACPubNumber
\end{flushleft}

\title{
{\large \bf \boldmath Measurement of \BDstarDss Branching Fractions 
and \BDstarDsstar Polarization 
with a Partial Reconstruction Technique}
}

%
\author{B.~Aubert}
\author{R.~Barate}
\author{D.~Boutigny}
\author{J.-M.~Gaillard}
\author{A.~Hicheur}
\author{Y.~Karyotakis}
\author{J.~P.~Lees}
\author{P.~Robbe}
\author{V.~Tisserand}
\author{A.~Zghiche}
\affiliation{Laboratoire de Physique des Particules, F-74941 Annecy-le-Vieux, France }
% ...
\collaboration{The \babar\ Collaboration}
\noaffiliation

\date{\today}

\begin{abstract}
We present a study of the decays \BDstarDss, 
using 20.8\invfb of \epem annihilation data recorded with the \babar\ detector.
The analysis is conducted with a partial reconstruction
technique, in which only the \Dsps and the soft pion from the $D^{*-}$ 
decay are reconstructed.
We measure the branching fractions 
${\cal B}(\BDstarDs) = (1.03 \pm 0.14 \pm 0.13 \pm 0.26) \%$
and 
${\cal B}(\BDstarDsstar) = (1.97 \pm 0.15 \pm 0.30 \pm 0.49) \%$,
where the first error is statistical, the second is systematic,
and the third is the error due to the \Dsphipi branching fraction uncertainty.
From the \BDstarDsstar angular distributions,  
we measure the fraction of longitudinal
polarization  
$\Gamma_L/\Gamma = (51.9\pm 5.0 \pm 2.8)\%$,
which is consistent with theoretical predictions based on factorization.

\end{abstract}

\maketitle

\section{INTRODUCTION}

Precise knowledge of the branching fractions of exclusive \B decay
modes provides a test of the factorization approach~\cite{ref:th-fact}.
Factorization neglects final state interactions between the 
quarks of the two final state mesons. 
The pattern of branching fractions for two-body \B decays to modes such as 
$D^{(*)}\pi$, $D^{(*)}\rho$~\cite{ref:Dstpifact} can be successfully accommodated 
in such a model.
However, it is possible that the factorization assumption is not applicable
to the decays $\B\to D^{(*)}X$, where the meson $X$ contains a heavy quark.
The current experimental uncertainties
for $\B\to\Dsps\Dstarb$ branching fractions~\cite{ref:cleo-dsinc}
do not allow us to perform a precise test of the factorization 
approach in this case. 

Further tests of factorization are provided by measuring the polarization in decays of \B mesons 
to vector-vector final states.  Within experimental errors, 
polarization measurements are consistent with 
factorization predictions for the final states $\Dstarb\rho$~\cite{ref:cleo-dstrho},
$\Dstarb\rho(1450)$~\cite{ref:dstrhoprime}, and $D_s^*\Dstarb$~\cite{ref:cleo-dds-polar}.

In this paper we present measurements of the branching
fractions$^1$\footnotetext[1]{Reference to a specific decay
channel or state also implies the charge conjugate decay or state. 
The notation \Dsps refers to either \Ds or \Dspstar.} ${\cal B}(\BDstarDss)$. 
We also report a measurement of the \Dspstar polarization in the decay \BDstarDsstar, 
obtained from an angular analysis. These results provide tests of factorization 
with increased precision.

\section{THE \babar\ DETECTOR AND DATA SET}

The data used in this analysis were collected with the \babar\
detector at the \pep2\ storage ring. An integrated luminosity of
20.8\invfb was recorded in 1999 and 2000 at the \FourS resonance,
corresponding to about 22.7 million produced \BB pairs.

A detailed description of the \babar\ detector is presented in
Ref.~\cite{ref:babar}. Only the components of the detector most
relevant to this analysis are briefly described here.
Charged particles are reconstructed with a five-layer, double-sided
silicon vertex tracker (SVT) and a 40-layer drift chamber (DCH) with a
helium-based gas mixture, placed in a 1.5 T solenoidal field produced
by a superconducting magnet. The charged particle resolution
is approximately $(\delta p_T/p_T)^2 = (0.0013 \, p_T)^2 +
(0.0045)^2$, where $p_T$ is the transverse momentum given in \gevc.  
The SVT, with a typical
single-hit resolution of 10\mum, provides measurement of the impact
parameters of charged particle tracks in both the plane transverse to
the beam direction and along the beam.
Charged particle types are identified from the ionization energy loss
(\dedx) measured in the DCH and SVT, and the Cherenkov radiation
detected in a ring imaging Cherenkov device (DIRC).  Photons are
identified by a CsI(Tl) electromagnetic calorimeter (EMC) with an
energy resolution $\sigma(E)/E = 0.023\cdot(E/{\rm GeV})^{-1/4}\oplus
0.019$.

\section{METHOD OF PARTIAL RECONSTRUCTION}
\label{sec:Analysis}

\par In selecting candidates for the decays \BDstarDss with $ D^{*-}
\rightarrow \Dzb \pi^- $, no attempt is made to reconstruct the
\Dzb decays.  Only the \Dsps and the soft $\pi^-$ from the $
D^{*-}$ decay are detected. In this way, the candidate selection efficiency is 
higher by almost an order of magnitude than that obtained with
full reconstruction of the final state. 
Given the four-momenta of the \Dsps and $\pi^-$, and assuming they
originate from a \BDstarDss decay, the four-momentum of the \Bz can
be calculated up to an unknown azimuthal angle $\phi$ around 
the \Dsps flight direction. This calculation uses the constraint of 
the known center-of-mass (CM) energy and the masses of 
the \Bz and $D^{*-}$ mesons. 
Energy and momentum conservation then allows a determination of the
four-momentum of the \Dzb, whose square yields the 
$\phi$-dependent missing mass
\begin{equation}
\mmiss = \sqrt { ( {\rm P}_B - {\rm P}_{\Dsps }- {\rm P}_{\pi})^2  } \ , 
\end{equation}
where ${\rm P}_B,\ {\rm P}_{\Dsps } \ {\rm and} \ {\rm P}_{\pi}$ 
are the four-momenta of the \Bz, \Dsps and the soft pion, respectively.
In this analysis the missing mass is defined with an arbitrary choice
for the angle $\phi$, such that the \Bz momentum ${\bf p}_B$ makes the
smallest possible angle with ${\bf p}_{\pi}$ and ${\bf p}_{\Dsps}$ in the 
CM frame.

\section{EVENT SELECTION}
\label{sec:selection}
For each event, we calculate the ratio of the second to the zeroth
order Fox-Wolfram moments, using 
all observed charged tracks and neutral clusters. This ratio is 
required to be less than 0.35 in order to 
suppress continuum $\epem \to \qqbar$ events, where $q = u,d,s,c$.

We reconstruct \Ds mesons in the decay modes \dstophipi,
\dstokstark and \dstoksk, with subsequent decays \phikk, $\Kstarzb\to
K^-\pi^+$ and $\KS\to\pi^+\pi^-$. These modes are selected since they
offer the best combination of large branching fraction, good detection efficiency,
and high signal-to-background ratio.  Charged tracks from the \Ds are required to
originate from within $\pm$10~cm along the beam direction and
$\pm$1.5~cm in the transverse plane, and leave at least 12 hits in the
DCH.

Kaons are identified using \dedx information from the SVT and DCH, as well as
the Cherenkov angle and the number of photons measured with the DIRC.
For each detector component $d = \{\rm SVT, \ DCH, \ DIRC\}$, a
likelihood $L^K_d$ ($L^\pi_d$) is calculated given the kaon (pion)
mass hypothesis.  A charged particle is classified as a ``loose'' kaon
if it satisfies $L^K_d / L^\pi_d > 1$ for at least one of the detector
components. A ``tight'' kaon classification is made if the condition 
$\prod_d L^K_d / L^\pi_d > 1$ is satisfied.

Three charged tracks consistent with originating from a common vertex are
combined to form a \Ds candidate. 

In the case of the decay \Dsphipi, two oppositely charged tracks must
be identified as kaons with both satisfying the loose criterion,
and at least one, the tight criterion.
No identification requirement is applied to the pion.  The
reconstructed invariant mass of the ${K^+K^-}$ candidates must be
within 8\mevcc of the nominal $\phi$ mass~\cite{ref:pdg}.  In the
decay \Dsphipi, the $\phi$ meson is polarized longitudinally,
resulting in the kaons having a $\cos^2\theta_{H}$ distribution, where
$\theta_{H}$ is the angle between the $K^+$ and \Ds directions in the $\phi$ rest
frame.  We require $|\cos\theta_{H}|>0.3$, which retains 97\% of the
signal while rejecting about 30\% of the background.

In the reconstruction of the \dstokstark mode, the $K^{-}\pi^{+}$
invariant mass is required to be within 65\mevcc of the nominal
$\Kstarzb$ mass~\cite{ref:pdg}. This wider window leads to 
a larger fraction of combinatorial background than in the \Dsphipi mode.  
To reduce this background, we require
$|\cos\theta_{H}|>0.5$. In addition, substantial background arises
from the decays $D^{+}\to \Kstarzb\pi^{+}$ and $D^{+}\to \Kzb\pi^{+}$,
which tend to peak around the nominal \Ds mass if the pion is misidentified
as a kaon. This background is suppressed by requiring that the kaon daughter of the $\Kstarzb$
satisfy the loose kaon identification criterion and the other
kaon, the tight criterion.

For the decay mode \dstoksk, with $\KS\to \pi^{+}\pi^{-}$, the
$\pi^+\pi^-$ invariant mass must be within 15\mevcc of the nominal
\KS\ mass~\cite{ref:pdg}, and the charged kaon must satisfy the tight
criterion. To improve the purity of the \KS sample, we require the
angle $\alpha$ between the \KS momentum vector and the \KS flight direction 
to satisfy $\cos\alpha>0.98$.

The invariant mass $M_{D_s}$ of \Dsp candidates is required to be
within three standard deviations ($\sigma_{D_s}$) of the signal
distribution peak $M_{D_s}^{\rm peak}$ seen in the data.

\Ds candidates satisfying these selection criteria are combined with
photon candidates to form \Dsgamma candidates. 
Candidate photons are required to satisfy $E_\gamma > 50$\mev,
where $E_{\gamma}$ is the photon energy in the laboratory frame, and
$E_\gamma^* > 110$\mev, where $E^*_{\gamma}$ is the photon energy in
the CM frame. When the photon candidate is combined with any other
photon candidate in the event, the pair must not form a good $\piz$
candidate, defined by a total CM energy $E_{\gamma\gamma}^* > 200$\mev
and an invariant mass $115 < M_{\gamma\gamma} < 155$\mevcc.

The \Dspstar candidates must satisfy $|\Delta M - \Delta M^{\rm peak}| < 2.5\,
\sigma_{\Delta M}$, where $\Delta M^{\rm peak}$ is the peak of the
signal $\Delta M=M(\Dsp\gamma)-M(\Dsp)$ distribution observed in the data.
The CM momentum of the \Dsps candidate
is required to be greater than 1.5\gevc.

\Dsps candidates are combined with $\pi^-$
candidates to form partially reconstructed \BDstarDss candidates.
Since the transverse momentum of the pion in signal events is less than 210\mevc,
these tracks are not required to have DCH hits.

Due to the high combinatorial background in the $\Delta M$
distribution, more than one $\Dspstar\pi^-$ candidate pair 
is found per event, with about a 20\% probability from signal Monte Carlo simulation. 
To select the best candidate in the
event, the following $\chi^2$ is calculated for each \Dspstar candidate
\begin{equation}
\begin{array}{c}
\chi^2 =
  \bigl[(M_{\inter}-M_{\inter}^{\rm peak})/\sigma_{\inter}\bigr]^2 + 
  \bigl[(M_{D_s}-M_{D_s}^{\rm peak})/\sigma_{D_s}\bigr]^2 \\ \\
  +\bigl[(\Delta M-\Delta M^{\rm peak})/\sigma_{\Delta M}\bigr]^2,
\end{array}
\end{equation}
where $M_{\inter}$ is the measured invariant mass of the intermediate 
$i=\phi$, $K^{*0}$, or \KS candidate, depending on the \Dsp decay mode, 
$M_{\inter}^{\rm peak}$ is the corresponding peak of the signal $M_{\inter}$ distribution, and
$\sigma_{\inter}$ is its width obtained from data. The candidate with the
smallest value of $\chi^2$ in the event is retained. 

\section{RESULTS}
\label{sec:results}

\begin{figure}
\begin{center}
        \includegraphics[width=0.48\textwidth]{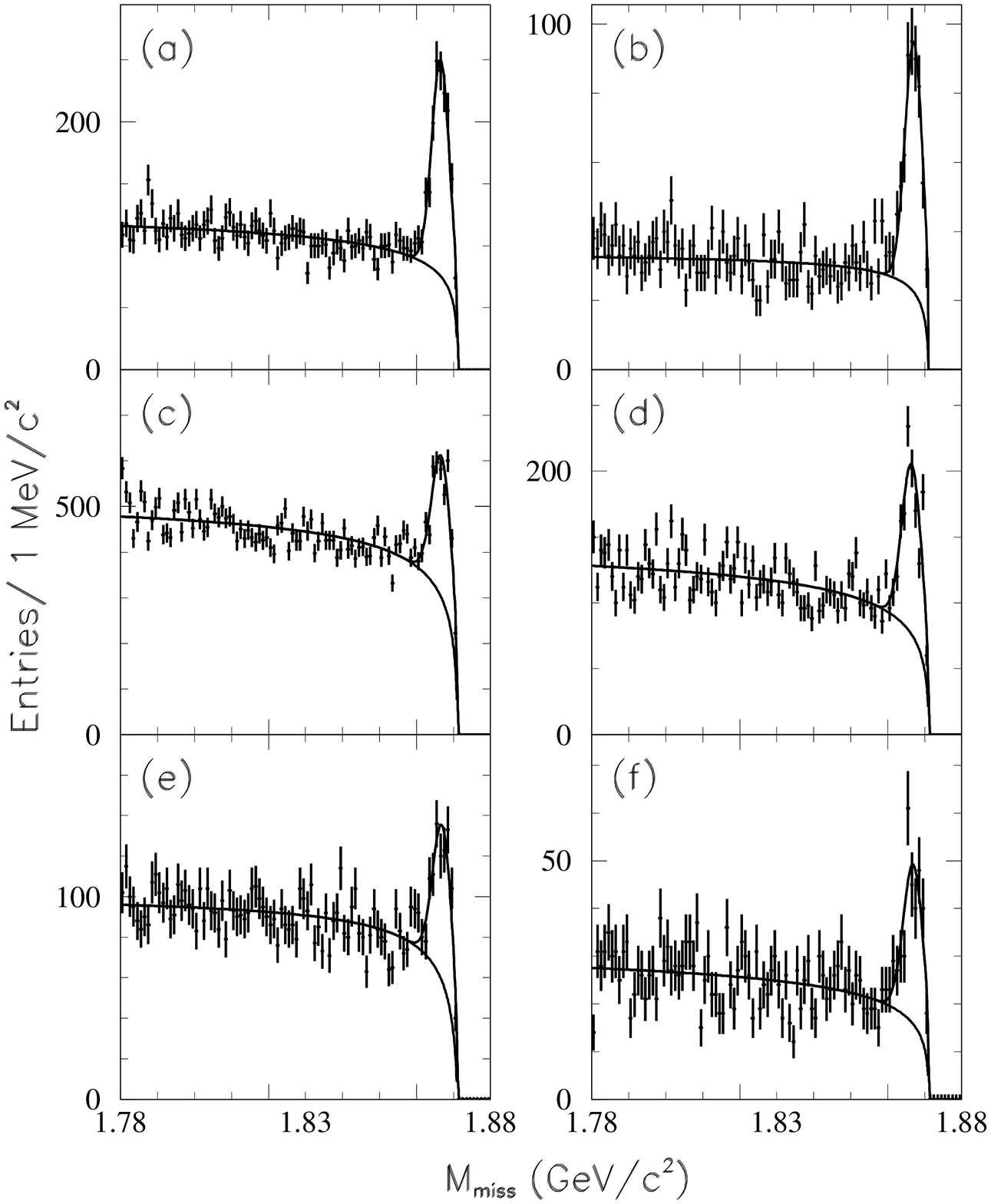} 
	\caption{Missing mass distributions of \B candidates in data.
		(a)~$\Dspi$ with \Dsphipi, 
		(b)~$\Dsstarpi$ with   \Dsphipi,
		(c)~$\Dspi$ with \dstokstark,
		(d)~$\Dsstarpi$ with \dstokstark,
		(e)~$\Dspi$ with \dstoksk, 
		(f)~$\Dsstarpi$ with \dstoksk.
		 The curves show the result of the fit (see text), 
		 indicating the signal 
		 and background contributions.}
	\label{fig:signal}
\end{center}
\end{figure}

The missing mass distributions of candidates for partially reconstructed \BDstarDss
decays are shown in Fig.~\ref{fig:signal}.  A clear signal peak is observed
in all modes. We perform a binned maximum likelihood fit to these
distributions. The fit function is the sum of a Gaussian distribution
and a background function given by
\begin{equation}
\label{math:bgr}
f_B(\mmiss) = \frac{C_1(M_0-\mmiss)^{C_2}\rule[-1.5mm]{0mm}{3mm}}{\rule[0mm]{0mm}{3mm}{C_3 + (M_0 - \mmiss)^{C_2}}} ,
\end{equation}
where $C_i$ are parameters determined by the fit, and $M_0 \equiv { M}_{
D^*}-{ M}_{\pi} = 1.871$\gevcc is the kinematic end point. The fits
find $3704\pm232$ and $1493\pm 95$ events under the Gaussian peak in the sum of the
$\Dsp\pi^-$ and $\Dspstar\pi^-$ distributions, respectively.
However, due to the presence of cross feed and self-cross feed, discussed below, 
further analysis is needed in order to extract the signal yields and 
the branching fractions.

We use a Monte Carlo simulation, which includes both \BB and
$q\bar{q}$ continuum events, to study the missing mass distributions
of the different background sources.  We consider two kinds of backgrounds: 
a peaking component that contributes predominantly at the end of the missing mass 
distribution in the signal region and a non-peaking component that is more uniform.
The non-peaking component is well modeled by the background function~(\ref{math:bgr}). 
The peaking component receives contributions from related channels due to
\begin{itemize}
\item
{\bf Cross Feed (CF):} if the soft photon from a \Dsgamma decay is not
reconstructed, \BDstarDsstar decays may lead to an enhancement under
the signal peak of the \Dspi missing mass spectrum.
Similarly, \BDstarDs decays may lead to a peaking enhancement in
the \Dsstarpi \mmiss spectrum, due to the combination of a \Ds with a random
photon.

\item
{\bf Self-Cross Feed (SCF):} this is due to true \BDstarDsstar decays
in which the \Ds is correctly reconstructed, but combined with a
random photon to produce the wrong \Dspstar candidate, resulting in a
peaking enhancement in the \Dsstarpi spectrum.
\end{itemize}
Table~\ref{tab:eff} presents the reconstruction efficiency of
correctly reconstructed signal \BDstarDss decays, as well as cross feed and
self-cross feed, found for simulated events in the signal region $\mmiss > 1.86$\gevcc.

\begin{table}
\begin{center}
\caption{ 
Efficiencies for \BDstarDss decay modes to contribute to the $\Dspi$ and $\Dsstarpi$
missing mass distributions in the signal
region $\mmiss > 1.86$\gevcc. Two different \BDstarDsstar Monte Carlo
samples have been used, one with longitudinal (long.) and the
other with transverse (transv.) polarization.}
\label{tab:eff}
\begin{tabular*}{0.48\textwidth}{lcc} \hline
& \multicolumn{2}{c}{Reconstructed mode} \\ 
Generated mode                               &  ~~~~~~~$\Dspi$~~~~~~~ & $\Dsstarpi$       	\\ \hline\hline
\BDstarDs			&  $(23.6\pm1.0)\%$	& $(1.7 \pm0.3)\%$	\\		\hline
\BDstarDsstar (long.)		&  $(9.0\pm0.3)\%$	& $(7.4 \pm0.3)\%$	\\
Self-Cross  Feed		&			& $(1.6 \pm0.1)\%$ 	\\		\hline
\BDstarDsstar (transv.)~~~	&  $(10.4\pm0.3)\%$	& $(6.9 \pm0.3)\%$	\\ 
Self-Cross  Feed		&			& $(1.4 \pm0.1)\%$	\\ 	\hline	\hline	
\end{tabular*}
\end{center}
\end{table}

In addition to these background sources, we also considered a possible
contribution from the charged and neutral \B decays $\B\to\Dsps
\Dbar^{**}$. These potential background sources were simulated with four $\Dbar^{**}$
states: the observed $\Dbar_1(2420)$ and $\Dstarb_2(2460)$ mesons,
and the $\Dstarb_0(j=1/2)$ and $\Dbar_1(j=1/2)$ mesons  predicted by HQET~\cite{ref:HQETDst}. 
Their contribution was determined to be
negligible, mainly due to the \Dsps CM momentum cut.
Multi-body decays $B\to\Dsps X$ are found not to contribute due to the same cut.

\begin{figure}
\begin{center}
        \includegraphics[width=0.48\textwidth]{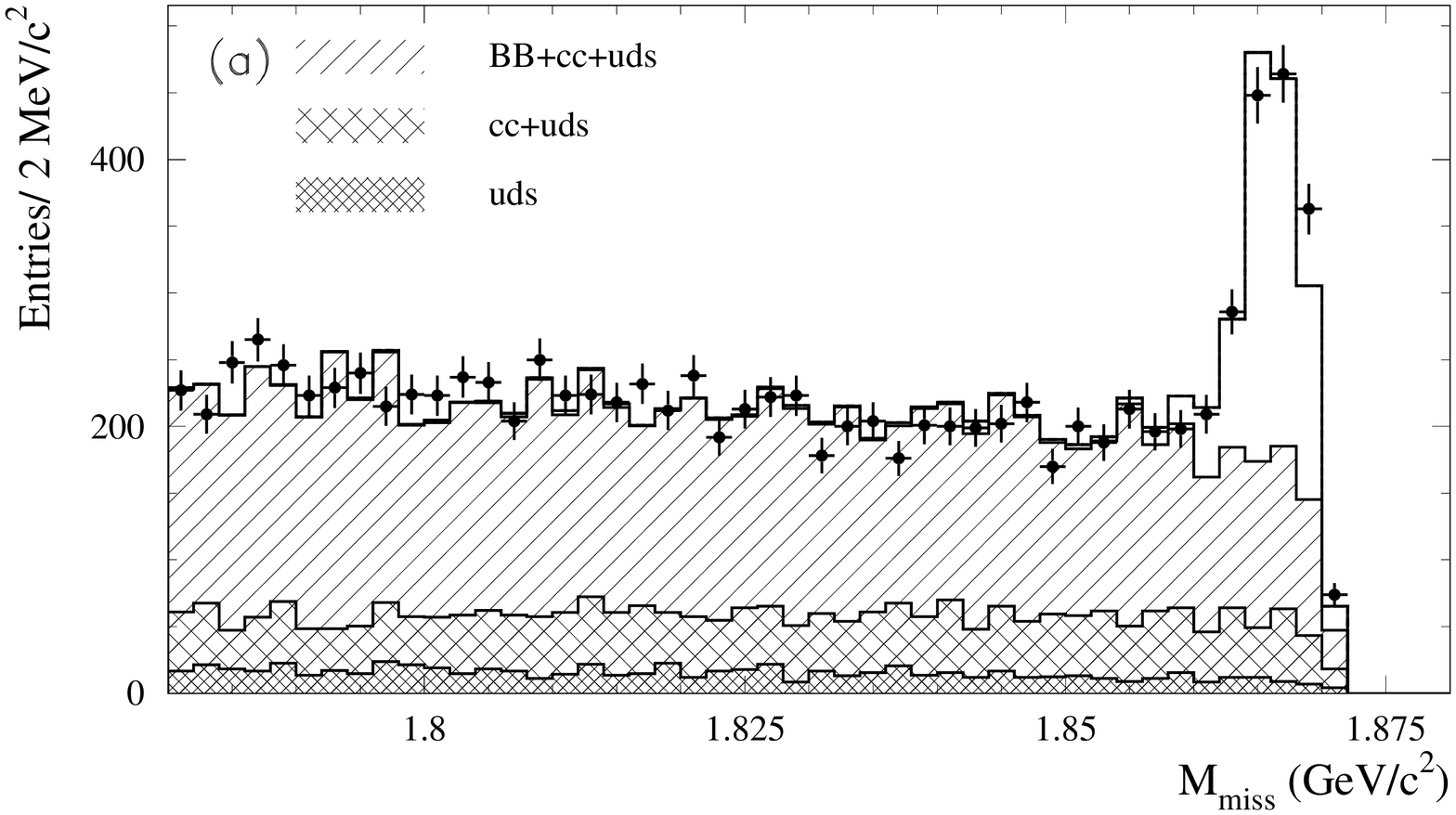} 
        \includegraphics[width=0.48\textwidth]{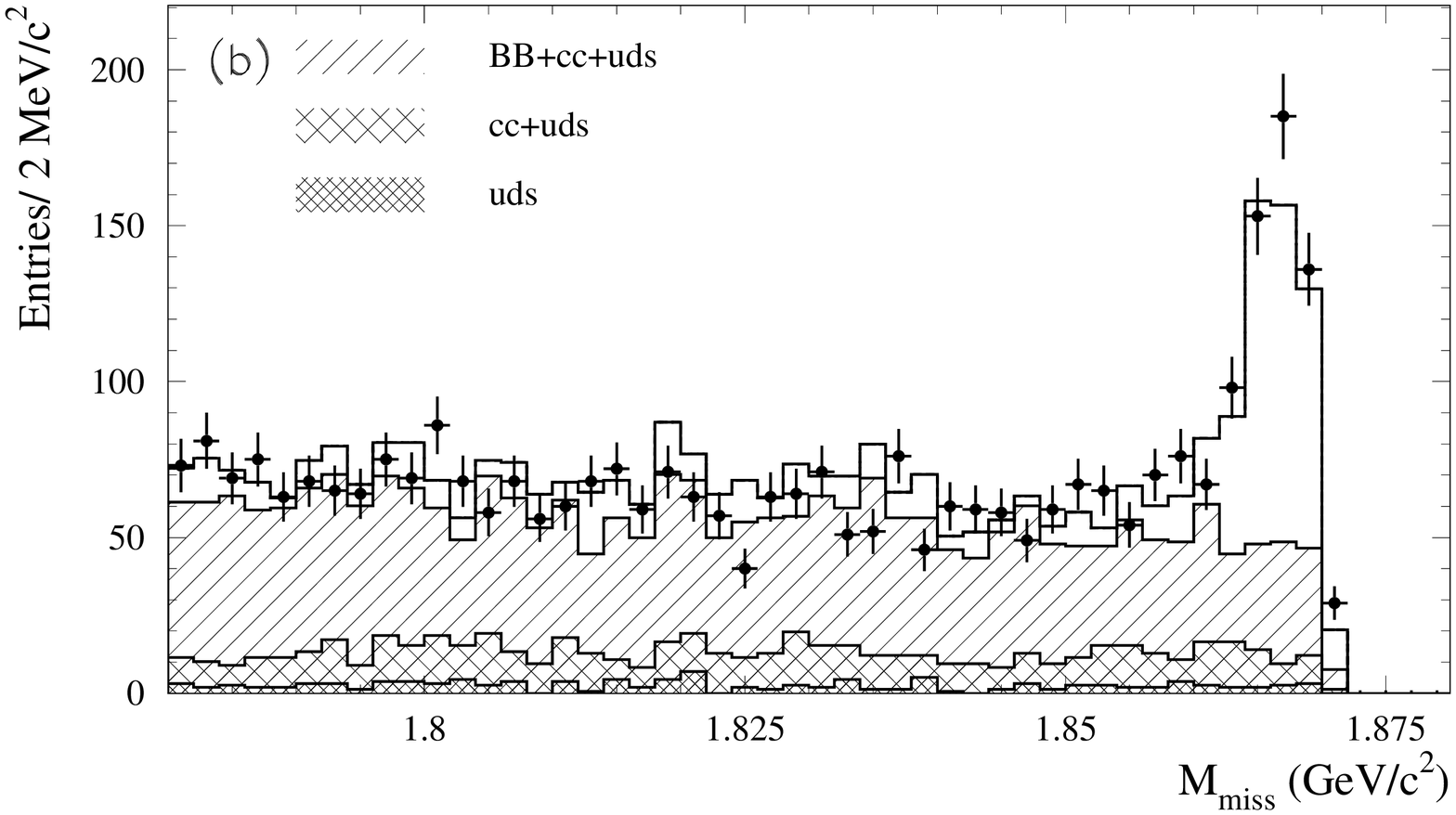} 
	\caption{Missing mass distribution for (a) $\Dspi$ and
		(b) $\Dsstarpi$ combinations
		for data (points with error bars) and Monte Carlo (histogram). 
		The contributions from the $\B\bar{\B}$, \ccbar and $q\bar{q}$ with 
		$q=u,\ d,\ s$ (labeled $uds$ in the figure) 
		are shown separately. The cross feed and self-cross feed backgrounds are included
		in the total histogram, but not in the hatched $\BB$ histogram.
		}
	\label{fig:mc}
\end{center}
\end{figure}

Figure~\ref{fig:mc} shows a comparison of the missing mass distributions
in data and Monte Carlo simulation. We assume 1.05\% and 1.59\% branching fractions 
for the \BDstarDs and \BDstarDsstar decays, respectively, in the Monte Carlo simulation.

The number of events in the peaks in the $\Dspi$ and $\Dsstarpi$ \mmiss distributions is
obtained from the fits described above.
The branching fractions are computed from these yields correcting 
for cross feed and self-cross feed background.
This is done by inverting the $2 \times 2 $ efficiency matrix, whose
diagonal elements correspond to the sum of signal and self-cross feed
efficiencies presented in Table~\ref{tab:eff}, and whose off-diagonal
terms are the cross-feed efficiencies.
The efficiencies corresponding to transverse and longitudinal
polarization of \BDstarDsstar have been weighted according to the
measured polarization discussed below.
With this procedure, the \BDstarDss branching fractions are determined
to be
\begin{equation}
\label{math:DsDst}
{\mathcal{B}} (\BDstarDs)  = 
(1.03 \pm 0.14 \pm 0.13  \pm 0.26 ) \%, 
\end{equation}
\begin{equation}
\label{math:DsstDst}
{\mathcal{B}} (\BDstarDsstar)  = 
(1.97 \pm 0.15 \pm 0.30  \pm 0.49   )\%,   
\end{equation}
and their sum is 
\begin{equation}
\label{math:brSum}
{\mathcal{B}} (\BDstarDss)  = 
(3.00 \pm 0.19  \pm 0.39 \pm 0.75 )\%,
\end{equation}
where the first error is statistical, the second is
the systematic error from all sources other than the uncertainty in
the \Dsphipi branching fraction, and the third error, which is 
dominant, is due the uncertainty in the \Dsphipi branching fraction
$\brphipi = (3.6\pm 0.9$)\%~\cite{ref:pdg}.
Correlations in the systematic errors between Eqs.~(\ref{math:DsDst}) and (\ref{math:DsstDst})
are taken into account in Eq.~(\ref{math:brSum}).
The sources of the systematic error are discussed in Sec.~\ref{sec:Systematics}.

The measurement of the fraction of the longitudinal polarization
$\Gamma_L/ \Gamma$ in the \BDstarDsstar decay mode is performed 
for candidates having missing mass in the signal region
($M_{miss}>1.86$\gevcc).
To minimize the systematic error due to large backgrounds, the
polarization measurement involves only the \Dsphipi channel, which
has the best signal to background ratio. Two angles are used:
the helicity angle $\theta_{\gamma}$ between the $D^{*-}$ and
the soft photon direction in the \Dspstar rest frame, and 
the helicity angle $\theta_{\pi}$ between the \Dspstar and 
the soft pion direction in the $D^{*-}$ rest frame.
Since the \B meson is not fully reconstructed, we compute
$\theta_{\gamma}$ and $\theta_{\pi}$ by constraining \mmiss to the
nominal $\Dz$ mass~\cite{ref:pdg} to obtain a unique 
solution for the azimuth $\phi$.

\begin{figure}
	\includegraphics[width=0.48\textwidth]{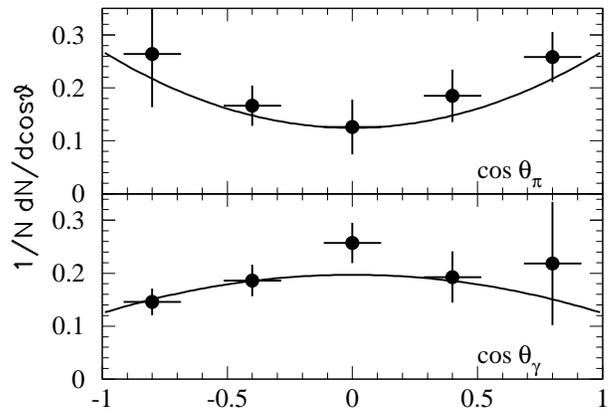}
\caption{Projections of the number of background-subtracted data events
on the $\cos\theta_\pi$ and $\cos\theta_\gamma$ axes.  The result of
the two-dimensional fit is overlaid.}
\label{fig:pol_proj}
\end{figure}

The two-dimensional distribution ($\cos\theta_\gamma$,
$\cos\theta_\pi$) is divided into five ranges in each dimension,
resulting in 25 bins.
The combinatorial background, as well as the cross feed and the self-cross 
feed obtained from Monte Carlo simulation, are subtracted
from this two-dimensional data distribution. The resulting signal
distribution is corrected bin-by-bin for detection efficiency, 
which is obtained from the simulation separately for each bin.
A two-dimensional binned minimum-$\chi^2$ fit is then performed to the
efficiency-corrected signal distribution with the fit function
\begin{eqnarray}
\frac{d^2\Gamma}{d\cos\theta_\pi\ d\cos\theta_\gamma } & \propto &  
\frac{\Gamma_L}{\Gamma}\ \cos^2\theta_\pi \sin^2\theta_\gamma+ \nonumber \\
&   & (1\ -\ \frac{\Gamma_L}{\Gamma}\ )  
	\sin^2\theta_\pi\frac{1+\cos^2\theta_\gamma}{4}.
\end{eqnarray}
The resulting fit has a $\chi^2$ of 23.1 for 25 bins with two floating
parameters ($\Gamma_L / \Gamma$ and total
normalization). Figure~\ref{fig:pol_proj} shows the data and the result
of the fit projected on the $\cos\theta_\gamma$ and $\cos\theta_\pi$
axes.

From the fit, the fraction of longitudinal polarization is determined to be
\begin{equation}
\label{math:GL}
\Gamma_L/\Gamma = (51.9 \pm 5.0   \pm 2.8 )  \%,
\end{equation}
where the first error is statistical and the second is systematic.

\section{Systematic Uncertainties}
\label{sec:Systematics}

The various contributions to the systematic errors on the branching fraction
and polarization measurements are summarized in Table~\ref{tab:syst}.  
The dominant systematic error for the branching fractions 
is the uncertainty on the three \Ds branching fractions.
To evaluate the uncertainty due to the background subtraction, 
the signal yield is determined using an alternative method
in which the number of events is extracted directly from 
the histogram after subtraction of the background, which is estimated 
with the Monte Carlo simulation.
The difference of the signal yields obtained in this way from the
results of the fit was taken as a systematic error.
This also accounts for the systematic error due to a possible deviation
of the signal shape from a Gaussian.

\begin{table}[!htb]
\caption{Sources of systematic uncertainties (\%) for the \BDstarDss 
branching fractions and \BDstarDsstar polarization measurements.}
\begin{center}
\begin{tabular}{lccc}
\hline
\hline
Source						&	$\Ds D^{*-}$\hspace{0.2cm}	&	$\Dspstar D^{*-}$\hspace{0.2cm}	&	$\sigma(\Gamma_L/\Gamma)$	\\ \hline
Background subtraction                          &                       &               &               \\
or modeling                                     &       2.7    		&	5.9	&	0.5	\\
Monte Carlo statistics                          &       4.2    		&	6.0	&	2.7	\\
Polarization uncertainty                        &       0.8		&	0.5    	&	-	\\ 
Cross feed					&	3.2		&      	2.4	&	-	\\ 
Number of \B pairs                              &       1.6    		&	1.6	&		-\\
$\BR(\phi\rightarrow { K^+K^-})$ 		&       1.6    		&	1.6	&		-\\
Particle identification                         &       1.0    		&	1.0	&	0.1	\\ 
Tracking efficiency                             &       3.6    		&	3.6	&	0.5	\\ 
Soft pion efficiency				&	2.0    		&	2.0	&	0.2	\\ 
Relative branching fractions			&       10.2		&	10.2	&	-	\\
$\BR(\Dsgamma)$  				& -			&	2.7	&		-\\
Photon efficiency                               & -      		&	1.3     &	0.1	\\
$\pi^0$ veto                                    & -      		&       2.7     &	0.3	\\ \hline 
Total systematic error				&  13.1    		&	15.1	& 	2.8	\\ \hline \hline 
\end{tabular}
\end{center}
\label{tab:syst}
\end{table}

The Monte Carlo statistical errors in the determination of the signal
and the cross feed efficiencies are included as a systematic error.
The uncertainty in the calculation of the \BDstarDsstar polarization is 
propagated to the branching fraction systematic error.
The systematic error due to the uncertainty on the efficiency 
for the reconstruction of charged particles 
is 1.2\% times the number of charged particles in the decay. An
additional error of 1.6\% is added in quadrature to account for the
uncertainty in the reconstruction efficiency of the soft pion.
The systematic error due to the \piz veto requirement was studied by measuring 
the relative \Dspstar yields in data and Monte Carlo with and without the \piz veto.

In the polarization measurement, the level of the various backgrounds
depends on the charged track, neutral cluster, and particle identification
efficiencies. The fit was repeated varying the background according to
the errors in these efficiencies, and the resulting variations in
$\Gamma_L / \Gamma$ were taken as the associated systematic error.

\begin{table}[hbt]
\caption{The fractional difference $\langle(N_{\rm D} - N_{\rm MC}) / N_{\rm MC}\rangle$,
averaged over all \mmiss bins, where $N_{\rm D}$ ($N_{\rm MC}$) is the
number of data (Monte Carlo) candidates in a given bin of the \mmiss
distribution of the given control sample. SB (SR) refers to the
$M_{D_s}$ or $\Delta M$ sideband (signal region) control sample.  WS
indicates wrong sign $\DsspiWS$ combinations, and $-p_{\Dsps}^*$
indicates that \mmiss was calculated from the negative of the \Dsps
CM momentum. The missing mass range $1.78 < \mmiss < 1.87$\gevcc is used 
for the control sample, except for the first line.}
\begin{center}
\begin{tabular}{lcc} \hline \hline
Sample type				& 	$\Dspi$		& $\Dsstarpi$	\\ 
\hline \hline
$1.78 < \mmiss < 1.85$\gevcc	& $-0.009\pm0.007$		& $\phantom{-}0.075\pm0.014$	\\ 
SB				& $-0.075\pm0.006$ 		& $\phantom{-}0.007\pm0.022$	\\
SR, WS	 			& $\phantom{-}0.006\pm0.008$	& $\phantom{-}0.044\pm0.015$	\\
SB, WS 				& $-0.060\pm0.007$		& $-0.008\pm0.024$	\\
SR, $-p_{\Dsps}^*$    		& $\phantom{-}0.015\pm0.009$	& $\phantom{-}0.075\pm0.016$	\\
SB, $-p_{\Dsps}^*$		& $-0.062\pm0.007$		& $-0.123\pm0.022$	\\ 
\hline
\hline
Average				& $-0.038\pm0.003$		& $\phantom{-}0.032\pm0.007$	\\
\hline \hline
\end{tabular}
\label{tab:mc_ratio}
\end{center}
\end{table}

To check that the simulation accurately reproduces the background \mmiss
distributions in the data, a thorough data-Monte Carlo comparison is
made in control samples containing no signal events. These samples are 
events with $1.78 < \mmiss < 1.85$\gevcc;
events in the \Dsp sideband $ 1.89 < M_{D_s} < 1.95$\gevcc or
$ 1.985 < M_{D_s} < 2.05$\gevcc;
events in the \Dspstar sideband $170 < \Delta M < 300$\mevcc;
wrong sign $\DsspiWS$ combinations in either the $M_{D_s}$ and $\Delta M$
sidebands or signal regions determined above;
and candidates in which \mmiss was calculated with the inverse of the 
\Dsps center-of-mass momentum $p_{\Dsps}^*$.
The comparison between data and Monte Carlo simulation for
these control samples is shown in Table~\ref{tab:mc_ratio}. 
The average level of discrepancy 
is used to estimate the uncertainty in the modeling of the background.

\section{SUMMARY}
\label{sec:Summary}

\par
In summary, based on a partial reconstruction technique, 
we have measured the branching fractions
\begin{eqnarray*}
{\mathcal{B}} (\BDstarDs)  & = & 
(1.03 \pm 0.14 \pm 0.13  \pm 0.26 ) \% \ , \; \\*[0.2 cm]
{\mathcal{B}} (\BDstarDsstar)  & = & 
(1.97 \pm 0.15 \pm 0.30  \pm 0.49   )\% \ , \;  \\*[0.2 cm]
{\mathcal{B}} (\BDstarDss)  & = & 
(3.00 \pm 0.19  \pm 0.39 \pm 0.75 )\%  \ .
\end{eqnarray*}
The fraction of the longitudinal \Dspstar polarization in
\BDstarDsstar decays is determined to be
$$ \Gamma_L/\Gamma = (51.9 \pm 5.0 \pm 2.8 ) \% .$$ 
This measurement is consistent with theoretical predictions assuming factorization,
which range from 50 to 55\%~\cite{ref:richman, ref:rosner}. Our 
results are also in good agreement with previous experimental
results~\cite{ref:cleo-dsinc, ref:cleo-dds-polar}.
\section{ACKNOWLEDGMENTS}
We are grateful for the excellent luminosity and machine conditions
provided by our \pep2\ colleagues, 
and for the substantial dedicated effort from
the computing organizations that support \babar.
The collaborating institutions wish to thank 
SLAC for its support and kind hospitality. 
This work is supported by
DOE
and NSF (USA),
NSERC (Canada),
IHEP (China),
CEA and
CNRS-IN2P3
(France),
BMBF and DFG
(Germany),
INFN (Italy),
NFR (Norway),
MIST (Russia), and
PPARC (United Kingdom). 
Individuals have received support from the 
A.~P.~Sloan Foundation, 
Research Corporation,
and Alexander von Humboldt Foundation.

\end{document}